\documentclass{elsart}

\usepackage{graphicx}

\usepackage{amssymb}

\newcommand{\AuAu}{Au+Au}
\newcommand{\pp}{\mbox{\textit{p}+\textit{p}}}

\newcommand{\pt}{\mbox{$p_T$}}

\newcommand{\gev}{\mbox{$\mathrm{GeV}$}}

\newcommand{\gevc}{\mbox{${\mathrm{GeV/}}c$}}

\newcommand{\RAA}{\mbox{$R_{AA}$}}
\newcommand{\vv}{\mbox{$v_{2}$}}

\newcommand{\nbin}{\mbox{$N_{\mathrm{bin}}$}}

\begin{document}

\begin{frontmatter}



\title{Identify bottom contribution in non-photonic electron spectra and \vv\ from \AuAu\ collisions at RHIC}


\vspace{-0.5cm}

\author{Yifei Zhang\corauthref{cor1}}

\corauth[cor1]{In collaboration with S.~Esumi, H.~Huang, Y.~Miake,
S.~Sakai and N.~Xu}

\address{Dept. of Modern Physics, University of Science and Technology of China, Hefei, Anhui, China, 230026}

\address{Lawrence Berkeley National Laboratory, 1 Cyclotron Road, MS70R319, Berkeley, CA, 94720}

\vspace{-0.6cm}

\begin{abstract}

We present a study on the spectra and elliptic flow \vv\ for heavy
flavor (charm and bottom) decayed electrons provided the relative
contributions of charm and bottom hadrons from the PYTHIA
calculations. We made a simultaneous fit to both measured
non-photonic electron spectra and \vv\ distributions. The results
suggest that the bottom contribution is not dominant for electron
$\pt<5$ \gevc\ in the 200 \gev\ \AuAu\ collisions.

\end{abstract}

\end{frontmatter}


Heavy quarks provide a unique tool to probe the partonic matter
created in relativistic heavy-ion collisions. In \AuAu\ collisions
at RHIC, a suppression of non-photonic electrons
spectra~\cite{STAREMCe,PHENIXe} and a non-zero non-photonic
electron \vv\ decreasing at $\pt>2$ \gevc~\cite{PHENIXv2} were
observed. In the measured non-photonic electron spectra and \vv\
distributions both contributions from charm-hadrons and
bottom-hadrons are mixed together. There are large uncertainties
in the model predictions for charm and bottom production in
high-energy nuclear collisions~\cite{FONLLvogt,Ko}. In order to
understand the heavy quark energy loss and flow, which are
sensitive to the properties of the hot/dense matter created at
RHIC energy, it is necessary to separate the charm from bottom
contributions. However, the isolation of charm and bottom
contributions with the semi-leptonic decay measurements alone
remains a challenge.

In this study, we assume the relative contributions of charm and
bottom from PYTHIA. We then fit the measured electron spectra
(\RAA) and \vv\ simultaneously. We find that the contribution of
bottom is not dominant for \pt\ (electron) $<5$ \gevc. This
electron \pt\ is corresponding to both D-meson and B-meson's \pt\
up to 5-7 \gevc. This simulation will provide a useful/necessary
constraint for heavy quark production model calculations.

\begin{figure}[htp]
\centering
\includegraphics[width=2.6in]{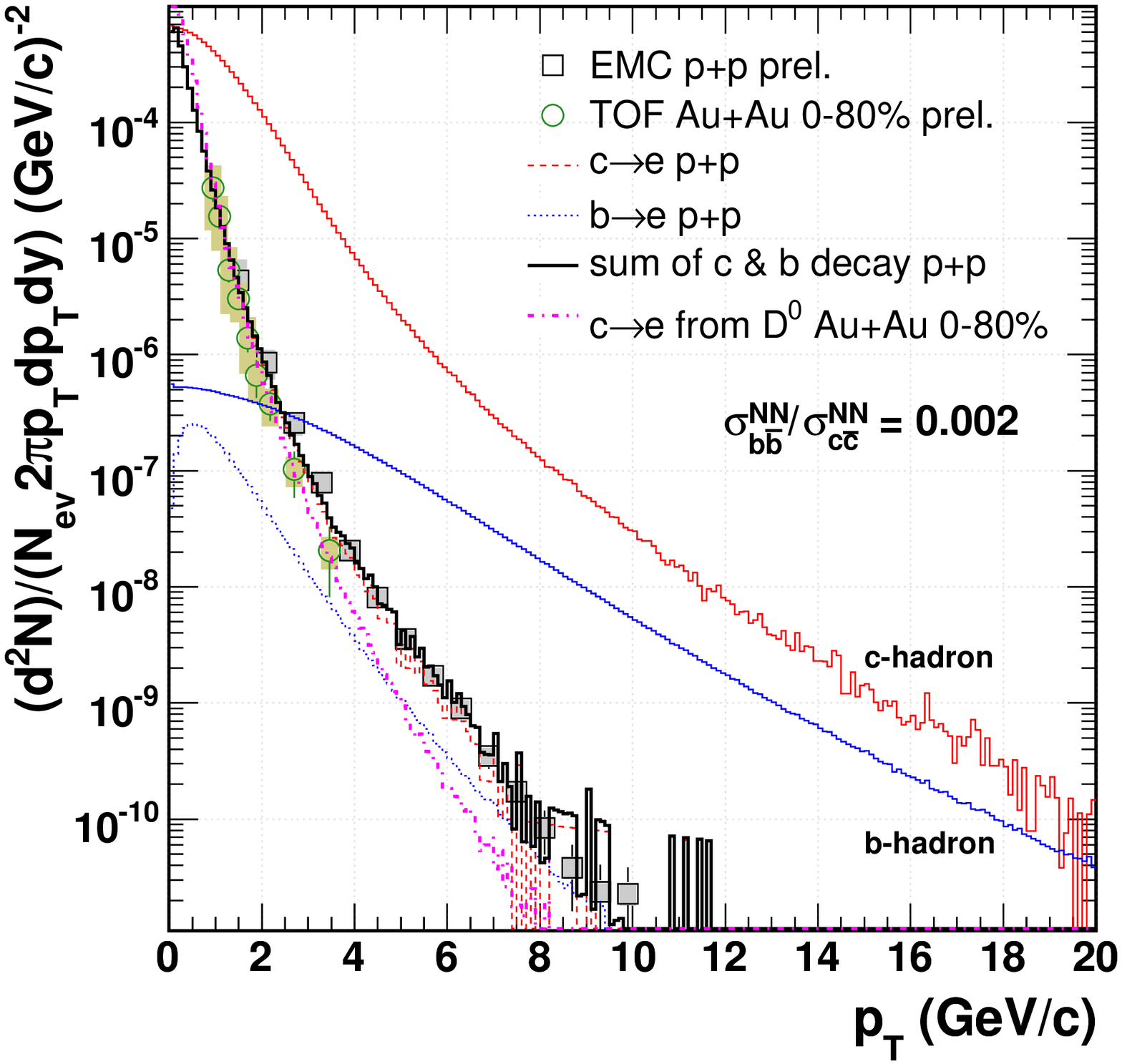}
\includegraphics[width=2.6in]{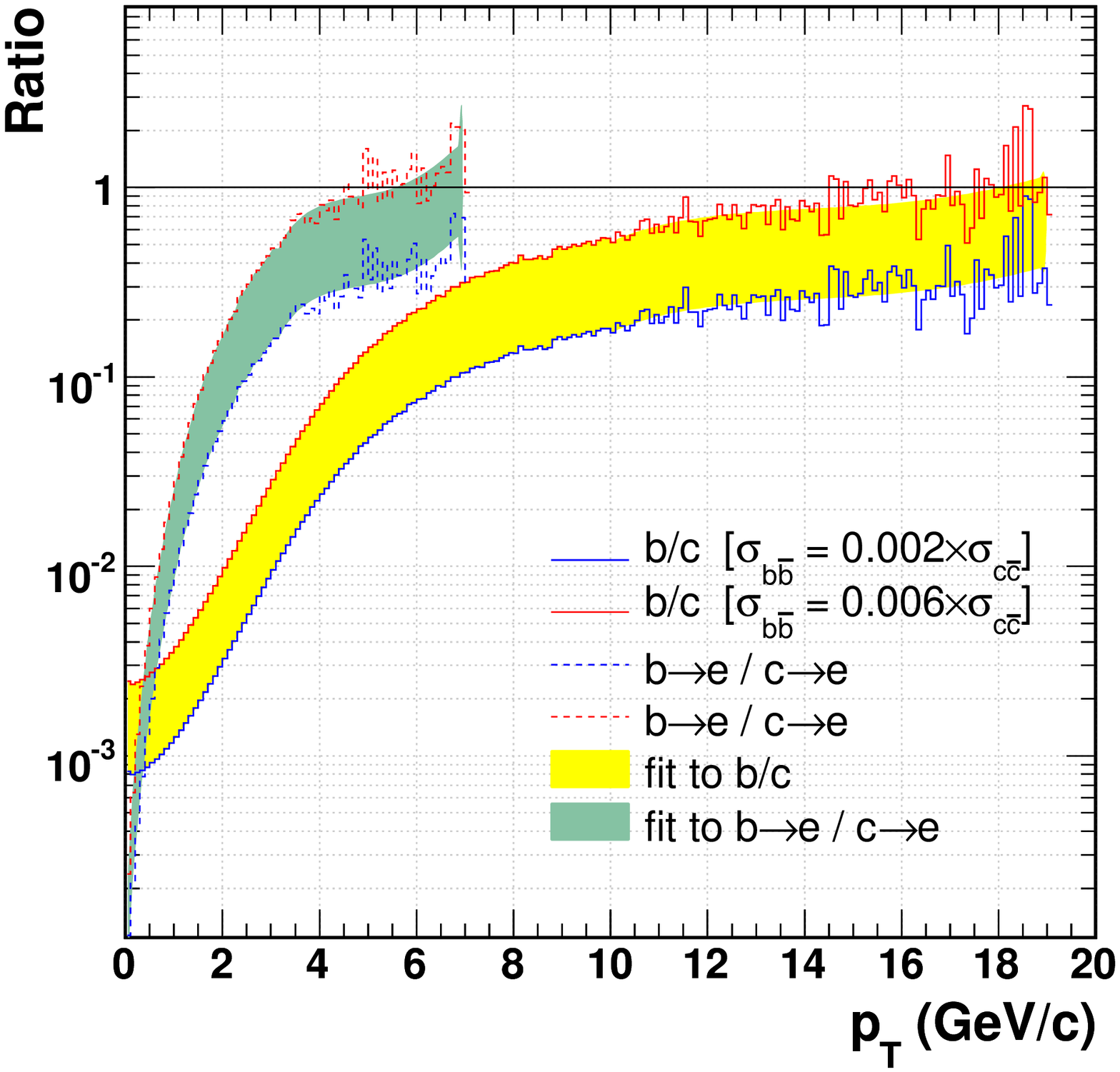}
\caption{Left panel: Charm/bottom-hadron and their decayed
electron spectra from simulation fit to experiment data in \pp\
and \AuAu\ collisions. Right panel: Spectra ratio bands are
according to $d\sigma_{b\bar{b}}^{NN}$/$d\sigma_{c\bar{c}}^{NN}$ =
0.002-0.006.}
\label{fig:Figure1}
\end{figure}

Since vector-meson decay form factor is used in PYTHIA, the
spectra from default PYTHIA parameters are soft~\cite{charmff}.
The heavy quark fragmentation function is modified from the
default Peterson function and the other parameter PARP(67) is also
tuned to describe the measured non-photonic electron spectra
~\cite{Xiaoyan}. The calculated charm-hadron spectrum is
normalized to $dN/dy$ = 0.028 measured in p+p
collisions~\cite{ppcharm} shown in the left panel of
Fig.~\ref{fig:Figure1}. A total cross-section per nucleon-nucleon
collision ratio of
$d\sigma_{b\bar{b}}^{NN}$/$d\sigma_{c\bar{c}}^{NN}$ = 0.002 is
assumed to normalize the bottom-hadron spectrum. Their decayed
electron spectra are shown as dashed and dotted curves
respectively. The total charm/bottom decayed electron spectrum
shown as thick solid curve is used to fit the non-photonic
electron spectrum from STAR EMC in \pp\ collisions
 (open squares) ~\cite{STAREMCe} by variating the bottom over charm cross-section
ratio. $c\rightarrow e$ from measured $D^0$ spectrum scale to
total charm (dot-dashed curve) via form factor
decay~\cite{charmff} and measured TOF electron spectrum (open
circles)~\cite{YifeiSQM06} in \AuAu\ 0-80\% are scaled by \nbin\ =
293. From the fit to the spectra, we can see that the agreement
between simulation and measurement is achieved when bottom
contribution is small. The right panel of Fig.~\ref{fig:Figure1}
shows the spectra ratio of bottom-hadron over charm-hadron (green
band) and their decayed electron (yellow band) as well. FONLL
calculation predicts the ratio from 0.0018 to 0.026 due to large
uncertainties of the cross-section~\cite{FONLLvogt}. Since
non-photonic electron spectrum is strongly suppressed in central
\AuAu\ collions compared the spectrum in \pp\ collisions, scaled
by number of binary collisions, bottom contribution seems small in
\AuAu\ collisions. So we assume
$d\sigma_{b\bar{b}}^{NN}$/$d\sigma_{c\bar{c}}^{NN}$ from 0.002 to
0.006 covered in the ratio bands. If charm is enhanced in RHIC
energy, this ratio will be smaller~\cite{YifeiSQM06}. With the
ratio of 0.002, we can see at electron $\pt\sim5$ \gevc, the
bottom decayed electron contribution is less than 20\% and
decrease fast when \pt\ goes lower.

In order to study on the bottom contribution to the non-photonic
electron \vv\ in \AuAu\ collisions, we propose two assumptions.
Firstly, we assume that
$d\sigma_{b\bar{b}}^{NN}$/$d\sigma_{c\bar{c}}^{NN}$ = $0.002 -
0.006$. The spectra ratios of charm-hadrons over bottom-hadrons
are the same in \pp\ and \AuAu\ collisions at RHIC. Charm-hadron
\vv\ is the same as the light hadron. Bottom-hadron \vv\ is the
same as the light hadron, or bottom-hadron \vv\ is zero.

\begin{figure}[htp]
\vspace{-0.4cm}
\centering
\includegraphics[width=4.5in]{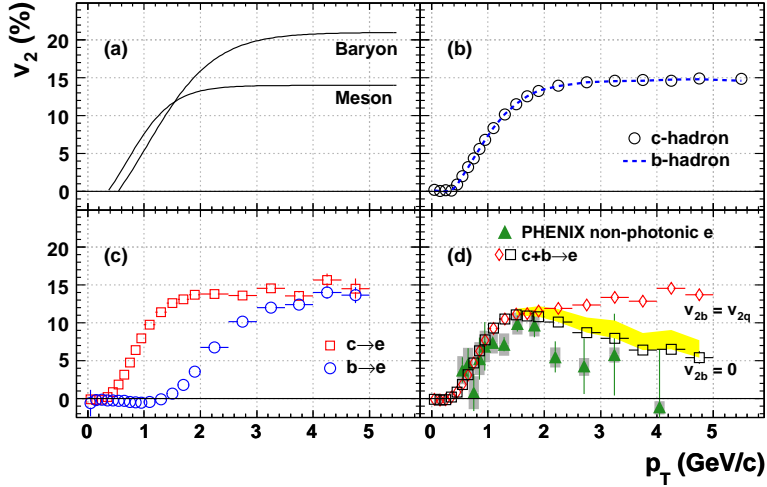}
\vspace{-0.25cm}
\caption{Panel (a): Light flavor meson \vv\ and
baryon \vv; Panel (b)(c): Assumed charm/bottom-hadron \vv\ and
their decayed electron \vv; Panel (d): Comparison between the
total decayed electron \vv\ and the PHENIX non-photonic electron
\vv.}
\label{fig:Figure2}
\end{figure}

Panel (a) of Fig.~\ref{fig:Figure2} shows the meson \vv\ and
baryon \vv\ from Ref.~\cite{XinPLB}. The assumed charm-hadron \vv\
and bottom-hadron \vv\ are shown in panel (b). Their decayed
electron \vv\ are shown in panel (c). The obvious mass effect of
the bottom decayed electron \vv\ pushing to high \pt\ is seen. The
total electron \vv\ from charm and bottom decay can be obtained by
applying the spectra ratio shown in the right panel of
Fig.~\ref{fig:Figure1}. If bottom-hadron \vv\ is assumed as the
dashed curve shown in panel (b), the total decayed electron \vv\
will be as the diamonds shown in the panel (d), which becomes flat
at higher \pt\ deviating from the PHENIX non-photonic electron
\vv\ (green triangles)~\cite{PHENIXv2}. If bottom-hadron \vv\ is
zero, the total decayed electron \vv\ will decrease as a function
of \pt, shown as the yellow band, due to the bottom contribution
according to the spectra ratio. The open squares are related to
the $d\sigma_{b\bar{b}}^{NN}$/$d\sigma_{c\bar{c}}^{NN}$ = 0.006.
Therefore, with this assumption, the decreasing of non-photonic
electron \vv\ is probably due to bottom contribution, but bottom
does not flow.

\begin{figure}[htp]
\vspace{-0.5cm}
\centering
\includegraphics[width=3.1in]{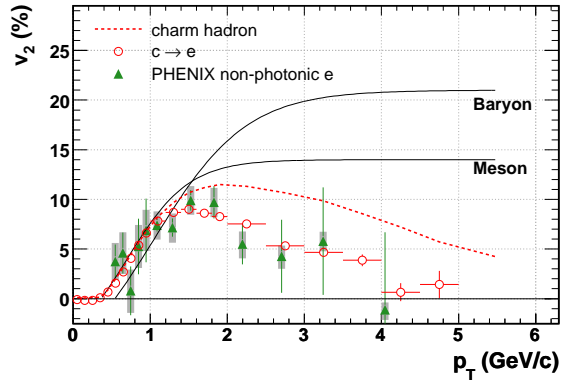}
\vspace{-0.25cm}
\caption{The decayed electron \vv\ from assumed
decreasing charm-hadron \vv\ fits to PHENIX non-photonic electron
\vv\ distributions. } \label{fig:Figure3}
\end{figure}

The second assumption is that bottom does not contribute. A
smaller and decreasing charm-hadron \vv\ is assumed shown as the
dashed curve in Fig.~\ref{fig:Figure3} at higher \pt. Its decayed
electron \vv\ (open squares) is used to fit the experimental data
(green triangles) by variating charm-hadrom \vv\ distribution. If
charm-hadron \vv\ is smaller than light hadrons and decreasing at
\pt$>2$ \gevc, its decayed electron \vv\ can describe the
experimental data.

In summary, charm/bottom and their decayed electron spectra and
\vv\ have been studied using PYTHIA simulation. If bottom
contribution is not dominant at \pt\ (electron) up to 5 \gevc,
both the measured non-photonic electron spectra (\RAA) and \vv\
can be described. This result should provide some constrain to
model calculations on heavy flavor productions though we reached
this conclusion with two hypotheses. The final solution to the
problem has to come from the direct meansurement of reconstructed
charm-hadron distributions.

We thank to the conference organizers. We would also like to
appreciate Drs. L.J.~Ruan and Z.B.~Xu for helpful discussions.



\end{document}